\documentclass[12pt]{article}
\usepackage{amsfonts}
\newcommand{\p}[1]{(\ref{#1})}

\newcommand{\be}{\begin{equation}}
\newcommand{\bea}{\begin{eqnarray}}
\newcommand{\ee}{\end{equation}}
\newcommand{\eea}{\end{eqnarray}}

\begin{document}
\thispagestyle{empty}
\topmargin-1cm
\renewcommand{\thefootnote}{\fnsymbol{footnote}}

\vspace{2cm}

\begin{center}
{\bf
On the SuperConformal Quantum Mechanics in the Nonlinear Realizations
Approach
}\vspace{0.3cm} \\

V.P. Akulov${}^{a,b}$ \footnote{E-mail:
akulov@juno.com}, Oktay Cebecioglu${}^{b,c}$\footnote{E-mail:
ocebecioglu@yahoo.com}
and
A. Pashnev${}^{d}$\footnote{E-mail:
pashnev@thsun1.jinr.dubna.su}\\
\vspace{0.3cm}

${}^a${\it Department of Natural Sciences, Baruch
College of the City University of New York}\\
{\it New York, NY 10010, USA}\\
\vspace{0.3cm}

${}^b${\it  Graduate School and
University Center,the City University
of New York,}\\
{\it 365 Fifth Ave, New York,NY 10016-4309,USA} \vspace{.3cm}\\

${}^c${\it Department of Physics,
Kocaeli University}\\
{\it Ataturk Bulvari, Anitpark Yanl,
Kocaeli, Turkey} \vspace{.3cm}\\

${}^d${\it Bogoliubov Laboratory of Theoretical Physics, JINR} \\
{\it Dubna, 141980, Russia}\vspace{.5cm}\\

{\bf Abstract}
\end{center}
\begin{center}
{\begin{minipage}{4.2truein}
                 \footnotesize
                 \parindent=0pt

In the framework of nonlinear realizations we rederive the action
of the $N=2$ SuperConformal Quantum Mechanics (SCQM). We propose also the
WZNW -- like construction of interaction term in the lagrangian with the help of
 Cartan's Omega forms.

\end{minipage}}\end{center}
                 \vskip 2em \par

\setcounter{footnote}0\setcounter{equation}0
\vfill
\setcounter{page}0
\renewcommand{\thefootnote}{\arabic{footnote}}

\newpage
\setcounter{equation}0
\section{Introduction}
The Conformal Quantum Mechanics (CQM)\cite{AFF} as well as its
supersymmetric generalization -- SCQM \cite{AP}-\cite{FR} are the simplest theories
for developing the methods of investigation of more complicated higher
dimensional field theories. The interest to the SCQM's with extended
supersymmetry \cite{P1}-\cite{IKL} is connected also with the fact that
these theories with anticommuting variables are exactly solvable not only on
the classical level \cite{AFF,AP,FR},\cite{AV,IKL1,IKL2}, but on the quantum
one as well, where superconformal group plays the role of dynamical symmetry
group. One should also note that in spite of its simplicity, the SCQM describes
the physical objects like a particle near horizons of black holes \cite{CDKKTP}
etc.\footnote{The extended SCQM is closely related with the Calogero model
with spin, which has many physical applications}

The geometrical meaning of CQM and SCQM can
be understood in the framework of nonlinear realizations of the symmetry
groups, underlying both theories - the group $SL(2,R)$ and its
supersymmetrical generalization $SU(1,1|1)$
respectively\cite{IKL1},\cite{IKL2}. In this approach the $N=4$ SCQM was
also constructed\cite{IKL2} (see also \cite{AIPT}). Some other variants of
$N=2$ and $N=4$ SCQM were analyzed in \cite{IKN},\cite{IKL}.
So, the nonlinear realizations method leads to the lagrangians, which may be
constructed in the framework of the usual superfield approach in which
supersymmetry is realized linearly in the standard manner.

In deriving of these results from the nonlinear realizations approach
the Cartan's Omega-forms technics is usually supplied by the so called
inverse Higgs effect \cite{IO}.  It is very powerful approach which
gives the possibility of covariant reduction of the number of the
variables by expressing some of them in terms of other. However, in
some cases the role of inverse Higgs effect can play the equations of motion
for some auxiliary variables like momenta in the Hamiltonian formulation
of the action integral. In general the question of interrelations
of these two approaches is not investigated yet.

One of the goals of the present paper is to show that
the consistent application of the nonlinear realizations approach gives the
possibility of constructing both the kinetic and interaction terms
for CQM and $N=2$ SCQM in the superfield approach without using the
inverse Higgs effect.

In Section 2 we reproduce results of \cite{IKL1} for CQM using the
natural matrix representation for the group $SL(2,R)$. In the framework of
nonlinear realizations method we construct invariant actions with the help
of Cartan's Omega-forms.  We give also the
method of construction of some additional invariant actions from the
Cartan's Omega-forms, though in this case we do not found new invariants - it reproduces
only old ones, constructed straightforwardly. These invariants plays the role
of the well known WZNW terms and are constructed as
 integrals over some additional
parameter $\sigma$ of Cartan's Omega-forms components.

After that we apply the same technics to $N=2$
SCQM using the matrix realization for the group $SU(1,1|1)$. We consider the
nonlinear realization of the coset $SU(1,1|1)/U(1)$, construct Cartan's Omega-forms
and show, how to build the kinetic part of the superfield action from the invariant coefficients
of these Omega-forms. The application of the method developed in the first
part of the paper gives us the possibility to construct from the Cartan's Omega-forms
the interaction part of the superfield action as well.
\setcounter{equation}0
\section{The Conformal Quantum mechanics}
\subsection{Conformal group and its matrix representation}
The conformal group in one dimensional space $SL(2,R)$ is a three-parameter subgroup of the
infinitedimensional reparametrization (diffeomorphisms) group on the line.
When the line is
parametrized by some parameter $s$
the generators of this
group are $L_m=is^{m+1}\frac{d}{d s}$ and
form the Virasoro algebra without central charge
\begin{equation}\label{Virasoro}
\left[L_n,L_m\right]=-i(n-m)L_{n+m}.
\end{equation}
If one restricts to the regular at the origin
$s=0$ transformations, it is convenient to parametrize
the group element as\footnote{The parameterizations
like \p{coset1} were firstly introduced in \cite{IK1,IK2} for 2-dimensional
(super)conformal groups.}
\begin{eqnarray}\label{coset1}
&&G=e^{i\tau L_{-1}} \cdot e^{ix_1L_1} \cdot e^{ix_2L_2} \cdot
e^{ix_3L_3}\ldots e^{i{x_0}L_0}.
\end{eqnarray}
The transformation laws of the coordinates in (\ref{coset1})
 under the infinitesimal
left action \begin{equation}\label{left1}
G'=(1+i\epsilon)G,\quad
\epsilon =
\epsilon^0 L_{-1}+{\epsilon^{1}}L_0+
{\epsilon^{2}}L_1+...+{\epsilon^{m}}L_m+... ,
\end{equation}
 are
\begin{eqnarray}\label{Transtau}
\delta \tau &=&\varepsilon(\tau)\equiv\epsilon^0+\epsilon^1\tau+
\epsilon^2\tau^2+\ldots\;,\\
\delta x_0&=&\dot{\varepsilon}(\tau),\label{Trans0}\\
\delta {x_1}&=&
-\dot{\varepsilon}(\tau)x_1+
\frac{1}{2}\ddot{\varepsilon}(\tau),\label{Trans1}\\
\delta {x_2}&=&
-2\dot{\varepsilon}(\tau)x_2+
\frac{1}{6}\stackrel{\ldots}{\varepsilon}(\tau),\label{Trans2}\\
&& \ldots \nonumber
\end{eqnarray}

In general the coordinate $x_n$ in (\ref{coset1}) transforms through the infinitesimal
transformation function ${\varepsilon}(\tau)$
and coordinates $x_k,\quad k\leq n$.
In addition the transformation law for parameter $x_n$ contains the term with $n+1$-st derivative of the
parameter ${\varepsilon}(\tau)$.

The simplest transformation law have the dimension-one coordinate
$\tau$ which transforms as the coordinate of the one-dimensional space under the
reparametrization. The coordinates $x_0$ and $x_1$ transform correspondingly as the dilaton
and one-dimensional Cristoffel symbol.

At this stage it is natural to consider all parameters as the fields
in one-dimensional space parametrized by the coordinate $\tau$. However, in general
all fields $x_m(\tau)$ can depend on some  coordinate $\sigma$,
which plays the role of additional parameter. Even more, such dependence on
the
additional parameter $\sigma$ can take place only for fields $x_m$ with
$m\geq M$ with fixed $M$. This will not lead to any contradictions with the
transformation laws \p{Transtau}-\p{Trans2}. As we will see, the introduction
of such additional coordinates gives the possibility to construct some
nontrivial invariants of the transformations \p{left1}.

As we already mentioned, the conformal group in one dimension is a subgroup
of \p{coset1}, namely the ones generated by $L_{-1}, L_0$ and $L_1$
\begin{equation}\label{coset2}
G_C=e^{i\tau L_{-1}} \cdot e^{ix_1L_1} \cdot e^{i{x_0}L_0}.
\end{equation}

As was shown in \cite{IKL1} the one-dimensional conformal mechanics
introduced in \cite{AFF} can be described on the language of
invariant differential Cartan's forms connected with the parametrization \p{coset2}
of the conformal group. Moreover, by the linear change of basis of the
conformal algebra one can describe\cite{IKN} on the same footing the ``new"
conformal mechanics of \cite{CDKKTP}.

In this Section we reproduce the results of \cite{IKL1} using the natural matrix realization
for the generators of $SL(2,R)$ group: translation $H=L_{-1}$, dilatation $D=L_0$ and conformal
transformation $K=L_1$
\begin{equation}\label{MatrixGenerators}
\begin{tabular}{c|ll|c|ll|c|lc|c}
    &$0$&$-1$&    &$0$ &$0$&                &$1$&$0$ &\\
$H=$&   &    &$K=$&    &   &$D=-\frac{i}{2}$&   &    &\\
    &$0$&$0$ &    &$-1$&$0$&                &$0$&$-1$&\\
\end{tabular}\\
\end{equation}
Such representation of the generators of the conformal group
can be easily generalized to the superconformal case\cite{VA}, including
the extended ones.

So, in the purely bosonic case the element of the conformal group in one dimension
can be parametrized as a product of three matrix multipliers
\begin{equation}\label{Kbose}
\begin{tabular}{c|ll|c|ll|c|ll|c}
  & $1$ & $it$&&$1$&$0$&&$x$&$0$&\\
$K_C=G_C=$&&&&&&&&&\\
  & $0$ &
$1$&&$-ix_1$&$1$&&$0$&$1/x$&\\
\end{tabular}\\
\begin{tabular}{c|ll|c|ll|c}
  & $1$ & $it$&&$x$&$0$&\\
$=$&&&&&&\\
  & $0$ &
$1$&&$ip$&$1/x$&\\
\end{tabular}\\
\end{equation}
The parameters in \p{coset2} and \p{Kbose} are connected by the relations
$t=-\tau,\; x=e^{x_0/2}$ and $p=-x_1 x$.  The conformal group transformation of these
new
variables are
\begin{equation}\label{CT}
  t'=\frac{at+b}{ct+d},\quad x'=\frac{x}{ct+d},\quad p'=(ct+d)p-cx,
\end{equation}
where parameters of the transformation are constrained by the unimodularity
condition $ad-bc=1$. Using the representation \p{coset1} one can calculate
also the transformations of functions
$x(\tau)$ and $p(\tau)$ under the most general (finite) reparametrization:
\begin{eqnarray}\label{TransGeneral}
t &\rightarrow&t'= f(t),\\
x(t)&\rightarrow&x'(t')=({\dot f}(t))^{1/2}x(t)\\
p(t)&\rightarrow&p'(t')=\frac{1}{({\dot f}(t))^{1/2}}p(t)+\frac{{\ddot f}(t)}{2({\dot f}(t))^{3/2}}x(t).
\end{eqnarray}

The invariant differential Cartan's form, calculated with the help of
\p{Kbose} is

\begin{equation}\label{Cbose}
\begin{tabular}{c|cc|c}
  & $(dx-pdt)/x$ & $idt/x^2$&\\
$\!\!\!\!\!\Omega_C=K_C^{-1}dK_C=$&&&\\
  & $i(x dp-p dx + p^2dt)$ &
$-(dx-pdt)/x$&\\
\end{tabular}\\ \!\!\!\!\!\!\!
\begin{tabular}{c|cc|c}
       & $\omega_{D}$ & $i\omega_{H}$&\\
$=\!\!$&               &               &\\
       & $i\omega_{K}$ &$-\omega_{D}$&\\
\end{tabular}\\
\end{equation}
All matrix elements in \p{Cbose} are invariant
under the transformations \p{CT}. One can recognize among them the einbein
differential form $\omega_{H}$.

\subsection{The action integral for Conformal Mechanics}

All these differential forms can be used for construction of an
invariant action. The simplest one is the linear combination
\begin{eqnarray}\label{BoseAction}
  S&=&-\frac{1}{2}\int \omega_{K}+
\alpha\int\omega_{D}-\Lambda\int\omega_{H}=\\
&&\int dt \left(-1/2(x {\dot p}-p{\dot x} + p^2)+\alpha({\dot
x}/x-p/x)-\frac{\Lambda}{x^2} \right)~.\nonumber
\end{eqnarray}
The first term in this expression is appropriately normalized to get the
correct kinetic term. The parameter $\Lambda$ plays the role of cosmological
constant.

One can find  $p$ by solving its equation of motion, insert it back in the
lagrangian and get the action of De Alfaro, Fubini and Furlan
\cite{AFF}
\be\label{ActionAFF}
S = {1\over 2} \int dt \left( \dot{x}^2 - {\gamma\over x^2} \right)~
\ee
with the coupling constant $\gamma=\Lambda+\alpha^2/2$. So, the parameter
$\alpha$ simply renormalizes the cosmological constant $\Lambda$.

At this point we should note that the integrand in the action $S$ \p{AFF} is
invariant only up to the total derivative, though it was deduced from the
expression \p{BoseAction} in which the integrand is strictly invariant
because it was constructed out of invariant Cartan's Omega forms
 $\omega_{K}$, $\omega_{D}$ and $\omega_{H}$. The reason of this lies
in the utilization of some equations of motion and partial integration
when \p{BoseAction} is transformed into \p{ActionAFF}. The same situation will
take place also in the more complicated case of $N=2$ SCQM.

The additional invariant actions can be constructed as \cite{AV1}
\begin{equation}\label{GenAct}
 S_F=\int\omega_{H}F(\frac{\omega_{K}}{\omega_{H}},\frac{\omega_{D}}{\omega_{H}})
\end{equation}
with arbitrary function $F$ of two invariant variables which are the coefficients
in the expressions of invariant differential one-forms $\omega_{K}$ and $\omega_{D}$ in terms
of only one (in dimension one) independent invariant one-form $\omega_{H}$.
 Some of these actions will
have form \p{ActionAFF}, but in general the actions \p{GenAct} will include
the higher degrees of the velocity $\dot{x}$.

The another sort of invariants in the action can be constructed by
introducing the dependence of the group element \p{coset2} or \p{Kbose}
on some parameter $\sigma$. Indeed, one can consider the special dependence
of the group element \p{Kbose} on some new parameter $\sigma$ such that
$t$ does not depend on it, whereas functions $x_0(t,\sigma)$ and
$x_1(t,\sigma)$ are subject to the following boundary conditions
\begin{equation}\label{BC}
 x_0(t,0)=x_1(t,0)=0 ,\quad
x_0(t,1)=x_0(t),\quad x_1(t,1)=x_1(t).
\end{equation}
So, the boundary group elements in \p{Kbose} are

\begin{equation}\label{Boundary0}
\begin{tabular}{c|ll|c|ll|c|ll|c}
  & $1$ & $it$&&$1$&$0$&&$1$&$0$&\\
$K_C(\sigma=0)=$&&&&&&&&&\\
  & $0$ &
$1$&&$0$&$1$&&$0$&$1$&\\
\end{tabular}\\
\begin{tabular}{c|ll|c}
  & $1$ & $it$&\\
$=$&&&$\cdot G_0$,\\
  & $0$ &
$1$&\\
\end{tabular}\\
\end{equation}
and
\begin{equation}\label{Boundary1}
K_C(\sigma=1)=K_C,
\end{equation}
where $G_0$ is the identity element of the group. The parameters $\epsilon^n$
(in our case $n=0,\;1,\;2$) of
transformation \p{left1} are assumed to be independent of $\sigma$. It means
that condition \p{Boundary1} will be the same for transformed quantities. On
the other hand the group element $G_0$ in the boundary condition
\p{Boundary0} will vary. But this variation is very simple, as one can see
from the transformation laws of $x_0$ and $x_1$. Moreover, both of them are
total derivatives.

All this can serve as an argumentation of the following construction,
leading in general to some integral invariants on the group.
In our case there are two independent invariant differential one-forms -
$\omega_{H}$ and $d\sigma$. So, the coefficients in expanding of all other
Cartan's forms in terms of these two forms are invariant as well. For
example, such one is
\begin{equation}\label{DInv}
 I=\frac{\omega_{D}}{d\sigma}=\frac{1}{x}\frac{d x}{d\sigma}
\end{equation}

In the presence of new coordinate $\sigma$ the invariant integration measure
is
\begin{equation}\label{Measure}
\int dv=\int \omega_{H} d\sigma=\int\frac{dt d\sigma}{x^2}.
\end{equation}
The result of integration over $\sigma$
\begin{equation}\label{S1}
  S_1= \int dv I=\int d t d\sigma \frac{1}{x^3}\frac{d
x}{d\sigma}=-\frac{1}{2}\int d t\frac{1}{x^2}|_{\sigma=1}+\frac{1}{2}\int d t\frac{1}{x^2}|_{\sigma=0}
\end{equation}
is the difference of two terms at points $\sigma=1$ and $\sigma=0$. The last
one is invariant by virtue of the transformation laws \p{Trans0}-\p{Trans1} of $x_0$ and $x_1$
near the identity element which corresponds to $x_0=0$ and $x_1=0$. Indeed,
they transform as total derivatives.
Because by construction $S_1$ is invariant, the term at the point $\sigma=1$ should be
also invariant, though the integrand in this term transforms as a total derivative.
 In the case under consideration it is
not wondering, because this invariant simply reproduces the already known
third term in \p{BoseAction}. Nevertheless, as we will see later, such
procedure of constructing can lead to new invariants in more complicated cases.

\setcounter{equation}0
\section{The $N=2$ SuperConformal Quantum mechanics}
\subsection{The matrix representation of the $N=2$ \\SuperConformal Group}
The $N=2$ SuperConformal group in one dimensional space is an eight-parameter subgroup of the
infinitedimensional $N=2$ Super Virasoro group with the following algebra of
its generators
\begin{eqnarray}\label{N2Virasoro}
  \left[L_n,L_m\right]&=&-i(n-m)L_{n+m},\\
\left[L_n,G_r\right]&=&-i(\frac{n}{2}-r)G_{n+r},\quad
\left[L_n,{\bar G}_r\right]=-i(\frac{n}{2}-r){\bar G}_{n+r},\\
\{G_r,{\bar G}_q\}&=&-2L_{r+q}-2(r-q)U_{r+q},\\
 \left[U_n,G_r\right]&=&-\frac{i}{2}G_{n+r},\quad
\left[U_n,{\bar G}_r\right]=\frac{i}{2}{\bar G}_{n+r}.
\end{eqnarray}
The indices $n,m$ are integer and $q,r$ - halfinteger ones. The $N=2$ SuperConformal
algebra contains in addition to the generators of Conformal algebra
(translation $H=L_{-1}$, dilatation $D=L_0$
and conformal transformation $K=L_1$) the $U(1)$ generator $U\equiv U_0$ and
generators of Poincar\`e ($Q=G_{-1/2}, {\bar Q}={\bar G}_{-1/2}$) and Conformal
($S=G_{1/2}, {\bar S}={\bar G}_{1/2}$) supersymmetries. All of these
generators can be realized in terms of $3\times 3$ graded matrices with vanishing supertrace
($Str M=M_{11}+M_{33}-M_{22}$):
\begin{equation}\label{MatrixGeneratorsN2B}
\begin{tabular}{c|ccc|c|ccc|c|ccc|c|ccc|c}
    & $0$ &$0$&$-1$& &$0$&0&$0$&                  &$1$&$0$&$0$&                &$1$&$0$&$0$&\\
$H=$&$0$&$0$&$0$&$K=$&$0$&$0$&$0$&$D=-\frac{i}{2}$&$0$&$0$&$0$&$U=-\frac{i}{2}$&$0$&$2$&$0$&\\
    & $0$ &$0$&$0$&  &$-1$&$0$&$0$&               &$0$&$0$&$-1$&               &$0$&$0$&$1$&\\
\end{tabular}\\
\end{equation}

\begin{equation}\label{MatrixGeneratorsN2F}
\begin{tabular}{c|ccc|c|ccc|c|ccc|c|ccc|c}
                        &$0$&$0$&$0$&           &$0$&$1$&$0$&    &$0$&$0$&$0$ &&$0$&$0$&$0$&\\
$\!\!\!\!\!\!Q=\sqrt{2}$&$0$&$0$&$1$&${\bar Q}=\sqrt{2}$&$0$&$0$&$0$&
$S=\sqrt{2}$&$i$&$0$&$0$ &${\bar S}=\sqrt{2}$&$0$&$0$&$0$&\\
    &$0$&$0$&$0$&    &$0$&$0$&$0$&                &$0$&$0$&$0$&    &$0$&$-i$&$0$&\\
\end{tabular}\\
\end{equation}

Our parametrization for coset space of $N=2$ SuperConformal group over the
$U(1)$ subgroup generated by $U$ is:
\begin{equation}\label{KN2}
\begin{tabular}{c|ccc|c|ccc|c|ccc|c}
  & $1$ &$\theta$& $it+\theta\bar\theta/2$&&$1$&$0$&$0$&&$x$&$0$&$0$&\\
$K_{SC}=\frac{G_{SC}}{U(1)}=$&$0$&$1$&$\bar\theta$&&$\psi$&$1$&$0$&&$0$&$1$&$0$\\
  & $0$ &$0$ &$1$&&$ix_1+\psi \bar\psi/2$&$\bar\psi$&$1$&&$0$&$0$&$1/x$&\\
\end{tabular}\\.
\end{equation}

The transformation laws of the coset space parameters under the infinitesimal left shift
\begin{equation}\label{leftN2}
K'_{SC}=(1+i\epsilon)K_{SC}=
\begin{tabular}{c|ccc|c}
  &$1+b/2 $&$\epsilon$&$i a$&\\
  &$\lambda$&$1$&${\bar \epsilon}$&\\
  & $i c$&${\bar\lambda}$&$1-b/2 $&\\
\end{tabular}\\ \cdot K_{SC}
\end{equation}are:
\begin{eqnarray}\label{TransformN2}
\delta t&=& a+bt+ct^2-\frac{i}{2}\epsilon\bar\theta -\frac{i}{2}\bar\epsilon\theta +
\frac{1}{2}(\lambda\theta-\bar\lambda\bar\theta)t,   \\
\delta \theta &=&\left(\frac{b}{2}+{c}t -
        \frac{1}{2}\bar\lambda\bar\theta\right)\theta +\epsilon -i\bar\lambda t    \\
\delta \psi &=&\left(-\frac{b}{2}-{c}t +
\frac{i}{2}c\theta\bar\theta+\bar\lambda\bar\theta\right)\psi +
\lambda-ic\bar\theta     \\
\delta x&=&\left(\frac{b}{2}+
{c}t-\frac{1}{2}\theta\lambda+\frac{1}{2}\bar\theta\bar\lambda \right)x    \\
\delta{(x_1)}&=&\left(-2bt+\bar\lambda\bar\theta+\theta\lambda
\right)x_1+b-\frac{i}{2}(\bar\lambda+ib\theta)\psi-
\frac{i}{2}(\lambda-ib\bar\theta)\bar\psi. \label{TransformN21}
\end{eqnarray}
One can see that in the point $(x=1, x_1=\psi=0)$ the variables $(x,
x_1, \psi)$ transform as a total derivatives.

The differential Cartan's form, calculated with the help of
\p{KN2} is

\begin{equation}\label{CN2}
\begin{tabular}{c|ccc|c}
  &$ \omega_D+i\omega_U  $&$\omega_Q$&$i\omega_H$&\\
$\!\!\!\!\!\Omega_C=K_C^{-1}dK_C=$&$\omega_S$&$2i\omega_U$&$\omega_{\bar Q}$&\\
  & $i\omega_K$&$\omega_{\bar S}$&$- \omega_D+i\omega_U  $&\\
\end{tabular}\\
\end{equation}
where
\begin{eqnarray}\label{OmegaN2}
  \omega _{D}&=&dx/x+\frac{1}{2}d\theta \psi
+\frac{1}{2}\bar{\psi}d\bar{\theta}-x_1dT\\
\omega _{Q}&=&(d\theta +i\bar{\psi}dT)/x\\
\omega _{\bar Q}&=&(d\bar{\theta}-i\psi dT)/x\\
\omega _{H}&=&dT/x^2\\
\omega _{S}&=&x\left(d\psi +x_1dT\psi +( ix_1+1/2\bar{\psi}\psi ) d\bar{
\theta}\right)\\
\omega _{\bar S}&=&x\left(d\bar{\psi}+x_1dT\bar{\psi}
+(-ix_1+1/2\bar{\psi}\psi )d\theta\right)\\
\omega _{K}&=&x^2\left(dx_1-\frac{i}{2}d\bar{\psi} \psi
+\frac{i}{2}\bar{\psi}d\psi+
x_1^2dT-x_1(d\theta\psi +\bar{\psi}d\bar{\theta})\right)\\
\omega_U& =&\frac{1}{2}\bar{\psi}\psi dT-\frac{i}{2}d\theta \psi +\frac{i}{2}\bar{\psi}d\bar{\theta}
\end{eqnarray}
and
\begin{equation}\label{dT}
  dT=dt-\frac{i}{2}d{\theta}\bar \theta
+\frac{i}{2}{\theta}d\bar\theta.
\end{equation}
Due to the fact that we consider the coset space $G_{SC}/U(1)$
instead of the whole group  $G_{SC}/U(1)$, not all of
Cartan's forms are invariant. Namely, $\omega _{Q}, \omega _{\bar Q}, \omega _{S}$
and $\omega _{\bar S}$ transform homogeneously as linear representations of
$U(1)$. As one can easily see
the forms $\omega _{Q}, \omega _{\bar S}$ carry the same charge under the $U(1)$ transformations,
whereas $ \omega _{\bar Q}$ and $\omega _{ S}$ carry the opposite equal charge.
 In turn, $\omega_U$ transform as a total differential.

\subsection{The action integral for $N=2$ SuperConformal Mechanics}
All Cartan's forms can be expanded in terms of three independent ones
 $\omega_{H}, \omega _{Q}, \omega _{\bar Q}$ using the formula
\begin{equation}\label{Diff}
 df(t,\theta,\bar\theta)=x\omega_QDf+x\omega_{\bar Q}\bar D
f+x^2\omega_H(\dot f-i\bar\psi Df+i\psi\bar D f),
\end{equation}
where $D$ and $\bar D$ are flat covariant derivatives
\begin{equation}\label{FlatDer}
  D=\frac{\partial}{\partial\theta}+
\frac{i}{2}\bar\theta\frac{\partial}{\partial t},\quad
\bar D=\frac{\partial}{\partial\bar\theta}+
\frac{i}{2}\theta\frac{\partial}{\partial t}.
\end{equation}
 The coefficients in such
expansions of $\omega _{H}, \omega _{D},\omega_S$ and $\omega _{K}$ are invariant
under the transformations \p{TransformN2}-\p{TransformN21} or transform as
some linear representation of $U(1)$, Two of these expansions which are useful in the construction
of invariants are
\begin{eqnarray}\label{ExpD}
 \omega _{D}&=&
\omega_{H}(xdx/dt-x_1x^2-i\bar{\psi}x Dx+i\psi x\bar{D}x)+\\
&&\omega _{Q}(Dx+x/2\psi
)+\bar{\omega}_{\bar{Q}}(\bar{D}x-x/2\bar{\psi}),\nonumber\\
\omega_S&=&x^2\omega_H(x\dot\psi-ix\bar\psi D\psi+ix\psi\bar D\psi)+\nonumber\\
&&x^2\omega_QD\psi+\omega_{\bar Q}(x^2\bar D \psi+ix_1x^2+1/2x^2\bar\psi\psi).
\end{eqnarray}
Since $ \omega _{D}$ and $ \omega _{H}$ are invariant, the coefficient
\begin{equation}\label{IDH}
I_{D/H}=  \frac{\omega _{D}}{\omega _{H}}=xdx/dt-x_1x^2-i\bar{\psi}x Dx+i\psi x\bar{D}x
\end{equation}
is invariant as well. At the same time the coefficients
\begin{equation}\label{IDQ}
 I_{D/Q}= \frac{\omega _{D}}{\omega _{Q}}=Dx+x/2\psi
\end{equation}
and
\begin{equation}\label{IDQB}
 -I_{D/\bar Q}=  - \frac{\omega _{D}}{\omega _{\bar Q}}=-\bar{D}x+x/2\bar{\psi}
\end{equation}
are mutually conjugated and transform with the opposite phase
under the transformations \p{TransformN2}-\p{TransformN21}.
So, their product $L_1=I_{D/Q}I_{D/\bar Q}$ as well as $L_2=I_{D/H}$ may be
used for construction of invariant lagrangians.

The coefficients
\begin{eqnarray}\label{invariants}
I_{S/\bar Q}=\frac{\omega_S}{\omega_{\bar Q}}&=&  x^2\bar D \psi+ix_1x^2+1/2x^2\bar\psi\psi,        \\
I_{\bar S/Q}=\frac{\omega_{\bar S}}{\omega_{Q}}&=& x^2 D \bar\psi-ix_1x^2+1/2x^2\bar\psi\psi
\end{eqnarray}
are mutually conjugated and inert
under the transformations \p{TransformN2}-\p{TransformN21}. So, their sum
$L_K=I_{S/\bar Q}+I_{\bar S/Q}$
gives one more possible term in the Lagrangian\footnote{The analogous construction
for $N=2$ Virasoro group gives the conformally invariant superfield description of
$N=2$ spinning particle\cite{P2}}. Indeed, it describes the
kinetic term of the SCQM in the superfield formulation.

 Using the invariant measure $dv=dtd\theta d\bar{\theta}$
one can construct the invariant action in the form
\begin{equation}\label{S0N2K}
 S_K=\int dtd\theta d\bar{\theta}\{I_{S/\bar Q}+I_{\bar S/Q}\}=
\int dtd\theta d\bar{\theta}\{ x^2\bar D \psi+x^2 D
\bar\psi+x^2\bar\psi\psi\}.
\end{equation}
With the help of the equations of motion for $\psi$ and $\bar\psi$
\begin{eqnarray}\label{Eqpsi}
&&-\bar{D}x+x/2\bar{\psi}=0,\\
&&Dx+x/2\psi=0,\label{Eqpsib}
\end{eqnarray}
the action $S_K$ can be rewritten
in the form
\begin{equation}\label{S0N2K1}
 S_K=4\int dtd\theta d\bar{\theta}\bar D xD x,
\end{equation}
in which the integrand is invariant only up to the total derivatives
(see the remark after the eq. \p{ActionAFF}).

Note, that the left hand sides of the equations \p{Eqpsi}-\p{Eqpsib} coincide
with the coefficients \p{IDQ}-\p{IDQB}. So, they vanish some part of the
Cartan's Omega form $\omega _{D}$, playing the role of inverse Higgs effect
\cite{IO}. It means that there is no need to use the eq's -\p{Eqpsi}-\p{Eqpsib}
as independent ones arising from the inverse Higgs effect as in \cite{AIPT}.

One can construct some other invariant terms for the action, for example
\begin{equation}\label{S0N2}
 S_0=\alpha\int dtd\theta d\bar{\theta}\{ (Dx+1/2x\psi) (\bar D x-1/2x\bar\psi)\},
\end{equation}
but this term leads simply to redefinition of overall coefficient in
\p{S0N2K1}.

The additional invariants in the action can be constructed by
the procedure described in the previous Section.
We again introduce the dependence of the group element \p{KN2}
on some parameter $\sigma$ such that
$t, \theta, \bar\theta$ do not depend on it, whereas functions
 $x_0(t,\sigma), x_1(t,\sigma)$ and$\psi(t,\sigma)$ are subject
to the boundary conditions
\begin{eqnarray}\label{BCN2}
&&  \ln x(t,0)=x_1(t,0)=\psi(t,0)=0,\\
&&\ln x(t,1)=\ln x(t),\; x_1(t,1)=x_1(t),\;
\psi(t,1)=\psi(t).\nonumber
\end{eqnarray}
So, the boundary group elements in \p{KN2} are

\begin{equation}\label{Boundary0N2}
\begin{tabular}{c|ccc|c|ccc|c}
  & $1$ &$\theta$& $it+\theta\bar\theta/2$&&$1$&$0$&$0$&\\
$K_{SC}(\sigma=0)=$&$0$&$1$&$\bar\theta$&&$0$&$1$&$0$&\\
  & $0$ &$0$ &$1$&                       &$0$&$0$&$1$&\\
\end{tabular}\\
\end{equation}
and
\begin{equation}\label{Boundary1N2}
K_{SC}(\sigma=1)=K_{SC}.
\end{equation}
Using the arguments of the previous Section one can show that the expression
\begin{equation}\label{S1N2T}
 \tilde S_1=\int dtd\theta d\bar{\theta}d\sigma\{\frac{\omega_D}{d\sigma}\}=
\int dtd\theta d\bar{\theta} d\sigma \frac{1}{x} \{\frac{dx}{d\sigma}\}
\end{equation}
is invariant and leads to additional invariant term in the action
\begin{equation}\label{S1N2}
 S_1=\int dtd\theta d\bar{\theta}\ln x
\end{equation}

So, the total action $S=S_0+S_1$ reproduces the action of
 $N=2$ SuperConformal Quantum mechanics \cite{AP}-\cite{FR} including both
the kinetic and potential terms.

\section{Conclusions}
In this paper, we applied the methods of nonlinear realizations approach for
construction of the actions of Conformal and $N=2$ SuperConformal Quantum
Mechanics. We have shown that both the kinetic and interaction terms
of these models can be constructed by using the invariant Cartan's
Omega-forms. The interaction part of the action looks like the well known WZNW
term. We have shown also that the Inverse Higgs Effect in both cases is a
consequence of the equations of motion for some variables. It would be
interesting to analyze the possibility of such duality between
 the Inverse Higgs Effect
and equation of motions for auxiliary variables in more complicated
theories, like $N=4$ SuperConformal Quantum
Mechanics. Besides, it is interesting to investigate in this model the
possibility getting with the help of the equations of motion of some
irreducibility conditions for the basic superfield, which were obtained
originally by the inverse Higgs effect \cite{IKL2}.

\section*{Acknowledgements}

A.P. thanks the members of Graduate School and
University Center, the City University
of New York, where the essential part of this work was
done and especially Prof. S. Catto for hospitality and
partial financial support. The work of A.P.  was partially supported by INTAS grant,
project No 00-00254.

\end{document}